\documentclass[prb]{revtex4}
\usepackage{epsf}
\begin{document}
\title{General Relativity and Compact Objects}
\author{Patrick Das Gupta}
\affiliation{Department of Physics and Astrophysics, University of Delhi, Delhi - 110 007 (India)}
\email{patrick@srb.org.in}
\begin{abstract}
Starting with the conceptual foundation of general relativity (GR) - equivalence principle, space-time geometry and special relativity, I train  cross hairs on  two characteristic  predictions of GR - black holes and gravitational waves. These two consequences of GR have played a significant role in relativistic astrophysics, e.g. compact X-ray sources, GRBs, quasars, blazars, coalescing binary pulsars, etc.

With quantum theory  wedded to GR, particle production from vacuum becomes a generic feature whenever event horizons are present. In this paper, I shall briefly discuss the fate of a `black hole atom' when Hawking radiation is taken into account.
In the context of gravitational waves, I shall  focus on the possible consequences of gravitational and electromagnetic radiation from highly magnetized and rapidly spinning white dwarfs. The discovery of RX J0648.0-4418 system - a WD in a binary with mass slightly over 1.2 $ M_{\odot}$, and rotating with  spin period as short as 13.2 s,  provides an impetus to revisit the problem of WD spin evolution due to energy loss.
\end{abstract} 
\maketitle
\section{INTRODUCTION}
In the centenary year of Einstein's general relativity (GR), when one takes stock  of  research studies concerning  accretion of matter and jet formation in the vicinity of  compact objects, one reaches an inescapable conclusion that the subject  has been accreting  relativistic  astrophysics pertaining to curved space-time, in an ever increasing manner. 

The three pillars of GR are - 
 (a) equivalence principle, (b) special relativity and (c) space-time geometry. 
 Because of (a), no matter how strong the  gravity is  in a given situation or how rapidly it varies with space-time, one is free to  choose a frame of reference of limited extent in space and time, such that gravity disappears in the frame (although  the gravitational tidal field does not).

 Such a local inertial frame (LIF) corresponds physically to a freely falling frame and mathematically to choosing a coordinate system  so that the metric tensor  is simply the Minkowski metric $\eta_{\mu \nu} $ all over this limited region. After marrying (a) and (b), GR insists that  mathematical forms of all non-gravitational physical laws in the LIFs take the same corresponding forms as they would  in `gravity free' inertial frames of special relativity. 
 
  From the above discussion, one can discern intuitively a link between gravity and space-time geometry - after all,  no matter how curved a 2-dimensional surface is, one can always choose a sufficiently  small patch $\Sigma $  on it such that the  distance between any two nearby points on $\Sigma $ can be obtained from $dl^2 = dx^2 + dy^2$ of Euclidean geometry. 
 
 To connect  LIFs at different space-time points, and to express physical laws in terms of arbitrary coordinates in reference frames of  size  as large   as one wishes, one needs the language of tensor calculus (or differential geometry) so that one  acquires an affine connection $\Gamma^\mu _{\alpha \beta}$ derivable from the metric tensor $g_{\mu \nu} $ and its derivatives. Although, the affine connection vanishes at a point in a LIF, its derivative does not. 
 
 Therefore,   true gravity  represented by the tidal gravitational field is related to the Riemann curvature tensor $R^\mu _{\nu \alpha \beta} $, a fourth rank tensor  constructed out of the connection $\Gamma^\mu _{\alpha \beta}$ and its derivatives. In mathematics, $R^\mu _{\nu \alpha \beta} $  determines whether the geometry is flat or curved.  This, in a sense,  completes the identification of gravity with geometry. While in the gauge theory framework, $\Gamma^\mu _{\alpha \beta}$ is analogous to   gauge potential with $R^\mu _{\nu \alpha \beta} $ as the corresponding gauge covariant field strength. 
 
 For the dynamics of bodies moving in pure gravity, the notion of gravitational mass becomes  superfluous in GR since particle  trajectories  are  geodesics of  space-time geometry determined from the line-element,
 $$ ds^2= g_{\mu \nu} dx^\mu dx^\nu \ .\eqno(1)$$
 Hence, it is not surprising that the world lines of freely falling test particles are independent of their inertial masses.
 
  On the other hand,  the dynamics of space-time geometry is determined by the Einstein equations,
 $$R_{\mu \nu} - \frac {1} {2} R g_{\mu \nu} = \frac {8 \pi G} {c^4} T_{\mu \nu}\eqno(2)$$
 where the Ricci tensor and Ricci scalar are  $R_{\mu \nu} \equiv R^\alpha  _{\mu \alpha \nu} $ and $R= g^{\mu \nu} R_{\mu \nu}$, respectively. $T_{\mu \nu} $ is the matter energy-momentum tensor whose various  components represent the flux of energy and momentum carried by matter in appropriate directions. When the gravity is weak and static, eq.(2) reduces to Newton's gravity,
 $$\nabla^2 \phi = 4 \pi G \rho \eqno(3)$$
 for a non-relativistic source with mass density $\rho $ and negligible pressure. The Newtonian gravitational potential $\phi $ is identified with the geometrical entity $(g_{00} - 1) c^2/2$.
 
  This, in a nutshell, is what GR is. The quest for  direct  detection of two of GR's cardinal predictions - black holes (BHs) and gravitational waves (GWs), is still on. I shall  briefly discuss elements of BH  physics in sections II and III. Towards the end, I will  describe some ongoing work of ours on GWs from rapidly rotating, magnetized  white dwarfs (WDs) in section IV.
 
\section{Cosmic Phenomena : General Relativity to the rescue} 
There is something inevitable about GR, at least in the classical domain. Apart from various experimental tests  that confirm the predictions of GR so far, there are multitude of astrophysical and cosmological observations which find natural explanations   only when GR is taken into consideration. 

For instance, the observed  redshifts in the case of extragalactic sources like quasars and radio-galaxies are unlikely to be Doppler shifts. This is obvious from the fact that a large fraction  of these sources have redshifts $1 \lesssim z \lesssim 7 $.  Doppler shift origin for such large values would imply  galactic size objects undergoing relativistic motion with very high Lorentz factor $\gamma \equiv ( 1 - v^2/c^2)^{-1/2}$, in which case, not only  would they be stripped off  their gaseous contents as they plough    through the intergalactic medium with speed $v \approx c$, but  would also have much greater observed mass $\gamma M_{gal}$, where  $M_{gal} $ is the galactic  mass in the rest frame.
Since, neither of these have any observational support, a much simpler explanation of such large redshifts is that of cosmological stretching of wavelengths due to  expansion of the universe, which ensues from GR solutions predicting  expanding (or  contracting) universe for homogeneous and isotropic matter distribution. 

Similarly, very large luminosities and rapid time variability associated with  active galactic nuclei (AGNs) can very effectively be explained in terms of accretion of matter close to the event horizon of a supermassive black hole (SMBH). Since $\xi^\mu = \delta^\mu_0$ is a time-like Killing vector for Schwarzschild space-time, a particle of rest mass $m$ lying at a radial coordinate $r$  from a Schwarzschild BH (SchBH) of mass $M$ has energy,
 $$E = \sqrt{ 1- R_s/r} \  m c^2 \eqno(4a)$$ where $R_s \equiv 2 G M/c^2 $ is the Schwarzschild radius of the BH. Therefore, by slowly lowering a test particle up to the event horizon of a BH, one can extract almost the entire rest energy $m c^2 $ of the particle. 
 
One may estimate from eq.(4a) the  maximum luminosity possible from matter slowly reaching upto a radius $r_0= 3 R_s$,
 $$L_{max} \cong \dot {m} c^2 \bigg [1- \sqrt {1 - \frac {R_s} {r_0}}\  \bigg ] \approx 0.18 \ \dot {m} c^2 \eqno(4b)$$
 where $\dot {m}$ is the rate of infall of matter.
 
 However, in accretion discs, matter is not at rest but  move in quasi-Keplerian orbits around the central BH. Hence, the above estimate needs to be modified to 
  illustrate in a simplified  manner why  BH  accretion is an efficient generator of energy. One may consider now a tiny volume element $dV$ of  an accretion disc, having rest mass $d m$,  orbiting around a SchBH.  Using the  Killing vector  $\xi^\mu = \delta^\mu_0$, one can obtain the energy $dE$ of the volume element    orbiting at a radial
coordinate $r$  with angular speed $d \phi /dt$,
$$dE = \frac { c^2 (1- R_s/r)\  dm} {\sqrt {1- R_s/r - (r/c)^2(d \phi /dt)^2}} \ . \eqno(5)$$

 As the innermost stable
circular orbit (ISCO)  around a SchBH has a radius,
$r_{ISCO}=3 R_s$ corresponding to an angular speed $\vert d \phi /dt  \vert = c/(3 \sqrt{6} R_s) $, one has 
  $ dE = 2 \sqrt{2} \ c^2 \ dm/ 3 $ from eq.(5),  implying a  radiative loss of energy $ c^2 \ dm - dE \cong  - 0.06 c^2 \ dm $, as the matter spirals in from very large $r$ to $r_{ISCO}$. Thus, accretion on to a SchBH can transform rest energy of matter into radiation with a maximum of $\sim 6 \% $ efficiency,   provided the viscous dissipation of kinetic energy
due to gradient of speed in the disc is efficient enough.  
Recent studies show that large magnetic fields threading through accretion discs can bring the ISCO closer to  the event horizon, and thereby possibly increase $L_{max}$ [1]. 
 
  Radiation from blazars and other AGNs display rapid fluctuations. High frequency radiation from blazars vary on
time scales $\Delta t \sim $ few hours [2].  Fast variations in luminosities find natural explanation in terms of accretion around SMBH. Causality arguments imply that transient processes taking place close to the event horizon of the SMBH can give rise to $\Delta t \sim \kappa \ r_{ISCO}/c$, where  $\kappa $ is a dimensionless parameter $\gtrsim 1$. Hence, observed $\Delta t \sim $1 hour time scale in a blazar  would mean $R_s \sim 3.6 \times 10^{14}/\kappa $ cm, implying a SMBH of mass $\sim 10^ 8 M_\odot $.
  
  As most astrophysical objects exhibit rotation, a BH formed out of the  collapse of a stellar system  is very likely  a Kerr BH. In addition to having an event horizon,  a Kerr BH  is also endowed with an ergosphere, a region  where  test particles cannot have constant spatial Boyer-Lindquist coordinates. Instead, the particles are  frame-dragged by the rotating BH because of the  Lense-Thirring effect.
   
Another interesting feature of the ergosphere is that the energy of a test particle in this region  can be negative, as measured by an inertial observer at infinity. This follows from  $\xi^\mu = \delta^\mu_0$ being a Killing vector also in the case of  Kerr geometry,  so that a freely falling particle having 4-momentum $p^\mu $ has energy $E = p^\mu \xi_\mu $ that is  conserved. In the ergosphere,  counter-rotating particles  can have $E < 0$.
 Penrose had proposed a mechanism to extract   rotational energy of a Kerr BH, wherein  a particle of energy $E_1$ decays  into two particles with energy $E_2$ and $E_3$ after it enters the ergosphere [3,4]. Since $E_1 = E_2 + E_3$,  the second particle escapes with energy $E_2 > E_1$ provided $E_3 < 0$. 

Clinching evidence for the existence of BHs are likely to ensue either from observation of  Penrose process in action around AGNs   and galactic BH X-ray sources or  from  detection of gravitational radiation due to quasi-normal modes at the time of BH formation during  the final  phase of a coalescing binary system.

Moving on to cosmology, late time acceleration of the universe deduced from SN Ia data as well as precision studies of cosmic microwave background radiation could  be caused either  due to a small value of the cosmological constant $\Lambda $ or by  a source of repulsive gravity. GR allows repulsive gravity provided there exists an exotic matter whose effective pressure is sufficiently negative.

 This can be easily illustrated by going beyond the Newtonian approximation of eq.(3). If one has a spherically symmetric object  with uniform density $\rho$ and pressure $p$ then it can be shown that the acceleration of a test particle near the surface of the object  is given by,
$$ \frac {d^2 \vec {r}} {dt^2} \cong  - \frac{4} {3} \pi G \ \bigg [ \rho + \frac {3  p}  {c^2} \bigg ]\  \vec {r} \eqno(6)$$
 Eq.(6) implies that if the object is made up of some kind of dark energy with $p < - \rho c^2/3 $, the resulting gravity is repulsive.
\section{Black holes and Quantum Theory}
In quantum field theory, the vacuum state $|0>$ is defined to be the lowest energy state, and therefore, corresponds to `no  particle' state. However,  variances of field operators in the vacuum state are, in general, non-zero. For instance, in    quantum electrodynamics, $<0|\hat{\vec{E}}. \hat{\vec{E}}|0>$ is non-zero signifying that the vacuum is seething with fluctuating electric field $\vec{E}$. The observed Lamb shift and Casimir effect are essentially due to this vacuum fluctuation of fields [5]. Another implication  is that vacuum is spontaneously, incessantly and randomly creating virtual pairs of particles of mass $m$  that last only  for short time scales $\sim \hbar/(2 m c^2)$, consistent with uncertainty principle.   

Now, according to eq.(4a), it costs very little energy to create a virtual pair at $r \approx R_s$, near the event horizon, so that if one of the pair falls into the BH the other can escape to infinity. This is a poor man's argument for Hawking radiation. But what Hawking had shown rigorously  was that the particles so produced, from the vicinity of a SchBH of mass $M$, have a thermal spectrum with a temperature [6],
$$T_H=\frac {\hbar c^3}{8 \pi G k_B M}= \frac {m^2_{Pl} c^2}{8 \pi k_B M} \cong 10^ {26} \ {}^\circ \mbox {K} \ \bigg ( \frac {M} {1\ \mbox{gm}} \bigg )^{-1} \eqno(7)$$
where $m_{Pl} = \sqrt{\hbar c/G}$  is the Planck mass. Eq.(7) demonstrates that BH thermodynamics is a common playground for general relativity, quantum theory as well as  statistical mechanics.

But according to quantum theory, states evolve in a unitary fashion (i.e. as prescribed by the Schrodinger equation) so that a pure state $|\psi >$ evolves into, in general, another pure state $|\chi >$.  Suppose, one forms a BH from the gravitational collapse of a system described by a pure state (e.g. a macroscopic Bose-Einstein condensate at zero temperature [7]). Since every phenomena is fundamentally a quantum phenomena, the BH formation too should be governed by a unitary evolution. In which case, even the final state of the BH along with its  entourage of Hawking radiation must correspond to a pure state.  

 On the contrary, Hawking radiation is thermal in nature, corresponding to a mixed state (i.e. random  distribution of pure states), and hence  cannot arise through a unitary evolution of an initially pure state.  A possible resolution is that the BH state is quantum mechanically entangled with the state of the radiated particles, so that the state of the combined system is pure. The apparent mixed state  arises because one  traces over the BH interior states which are inaccessible to an observer outside the event horizon.  But, what happens when the BH evaporates completely? For, then there is only radiation and no event horizon to form a shroud hiding the BH states.  This is the so called BH information loss problem, which in recent times is causing novel  ideas to emerge, ranging from fuzz ball to fire wall  [8,9].

Interestingly enough, I will show using simple arguments that Hawking radiation prevents a BH to capture an elementary particle into a quantum mechanical bound state. Using Bohr model for a gravitational bound state consisting of a BH of mass $M$ and a test  particle of mass $m$, one obtains the quantized energy levels (however, for more accurate energy levels, see [10]),
$$E_n = -\ \frac {G^2 M^2 m^3} {2\ n^2 \hbar^2} = -\ \frac {G\ M\ m} {2\ r_n} \eqno(8a)$$
corresponding to  orbital radii,
$$r_n = \frac {n^2 \hbar^2} {G\ M\ m^2}\ \ \ \mbox{with}\ \  n=1,2,...\eqno(8b)$$
Now, for a stable bound state, it is necessary that the closest orbital radius satisfies $r_1 > R_s$, as otherwise the particle will simply be swallowed by the BH. This implies,
$$ M < \frac {m^2_{Pl}} {\sqrt{2}\ m} \eqno(9a)$$
so that from eq.(7) the corresponding Hawking temperature is,
$$T_H > \frac {\sqrt{2}\  m c^2}{8 \pi k_B}\ .\eqno(9b)$$
From eqs.(8a) and (8b), the binding energy  $E_B$ of such a system is,
$$E_B= \frac {G\ M\ m} { 2\ r_1} < \frac {G\ M\ m} {2\ R_s} < \frac {1} {4} m c^2 \ .\eqno(9c)$$
While, from eqs.(9b) and (9c), the typical thermal energy of a particle emitted from the vicinity of the event horizon due to Hawking process is $\sim k_B T_H > \sqrt{2}\  m c^2/8 \pi $, which  is $\gtrsim E_B$. Therefore, Hawking radiation is destined to break up such a gravitationally bound `black hole atom'. 
\section{Gravitational Radiation from White Dwarfs}
Surface magnetic fields in white dwarfs (WDs) vary over a wide range. Some WDs  have  fields as high
 as $10^9 \ G$ [11-13]. Fraction of isolated WDs with magnetic field greater than $10^6 \ G$ is about 
 $10\ \%$. Mechanical stress induced by large interior magnetic field can make a spinning WD non-axisymmetric, and thereby   lead to emission  of GWs due to its changing mass quadrupole moment [14]. 
 
   WDs in close binaries tend to spin at faster rates. For instance, the WD in AE Aquarii has a spin period of only 33 s and  a spin down rate of  $ \dot {P}= (5.64 \pm 0.02)\times 10^{-14} \ s\ s^{-1}$, implying a magnetic dipole moment   $\mu \approx 1.5 \times 10^{34} \ \mbox {G cm}^3$. Its polar magnetic field is likely to be about $10^8$  G [15]. As of now, the fastest spinning WD is associated with the binary system  RX J0648.0-4418 having  spin period of  $13.18 \  s$ and mass $1.28 \pm 0.05 \ M_\odot $ [16-18]. Detection of GWs  from such rapidly spinning WDs with space based instruments like LISA could shed light on their interior magnetic fields.

The luminosity due to magnetic dipole radiation from a compact magnetized source that is spinning with an angular speed  $\Omega $ and located at a distance $d$ from us is given by $L_{EM}= (2\ \mu^2 \ \sin^2 \alpha \ \Omega^4)/3\ c^3 $,  
where $\mu=\frac{1}{2} B_p R^3 $ is the magnetic dipole moment of the object and $\alpha $ is the angle between the spin axis and  the magnetic dipole. 

On the other hand, the GW amplitudes from the  rotating WD   are [14]:  
$ h_\oplus=  h_0 \sin^2 \alpha\ \cos [2\Omega (t-t_0)] $
and 
  $h_\otimes=  h_0 \sin^2 \alpha \ \sin [2\Omega (t-t_0)] $ 
where,
 $$h_0\equiv - \frac {6 G\ \not I_{zz} \Omega^2 } {c^4 \ d} $$
  and the reduced mass quadrupole moment,
 $$\not I_{zz} \equiv \int{\rho (z^2-\frac {1}{3}r^2)d^3r}= - \beta_6 \ \delta_M \ M \ R^2 $$ 
 with  parameter $\delta_M $  being  defined as the ratio of magnetic energy to gravitational potential energy,
$$\delta_M = \frac{\int {(B^2/8\pi) d^3r}}{(\alpha_3 G M^2/R)} \approx  \frac {R^4 <B^2>}{6 \alpha_3 G M^2} \ .$$
The reduced quadrupole moment $\not I_{zz}$ has been obtained assuming the z-axis to be along the magnetic dipole moment. The parameters $\alpha_3$ and $\beta_6$ are of order unity whose values depend on the  mass density profile.

Then, the GW energy flux is given by, 
$$F_{GW}=\frac{c^3} {16 \pi G} [ {\dot {h}_\oplus}^2 + {\dot {h}_\otimes}^2 ] = \frac{c^3} {4 \pi G} h^2_0 \  \Omega^2\ \sin^4 \alpha $$
from which one can estimate the GW luminosity to be,
$$L_{GW} \approx 4 \pi d^2 F_{GW}= \bigg (\frac {\beta_6} {\alpha_3} \bigg )^2 \frac {<B^2>^2 R^{12} \ \Omega^6 \sin^4 \alpha} {G M^2 c^5}$$
Therefore, the  ratio of GW to EM power is given by [19],  
$$\frac {L_{GW}} {L_{EM}}= 6 \bigg (\frac {\beta_6} {\alpha_3} \bigg )^2 \frac {<B^2>^2}{B_p^2}\ \ \ \frac {R^6 \ \Omega^2 \sin^2 \alpha} {G M^2 c^2}\eqno(10a)$$
$$=37.5 \bigg (\frac {\beta_6} {\beta_I \alpha_3} \bigg )^2 \frac {<B^2>^2}{B_p^2}\ \ \ \frac {R^2 \ J^2 \sin^2 \alpha} {G M^4 c^2}\eqno(10b)$$
where $J=0.4 \beta_I \ M\ R^2 \Omega $ is the WD's spin angular momentum with $ \beta_I$ being a parameter of order unity whose value depends on the mass distribution. 

For AE Aquarii, a cataclysmic variable [15]:
 $ R=10^{8.8} $ cm, $ B_p\approx 10^8 $ G,  $ P= 33$ s, $M=0.65 \ M_\odot - 1.2 \ M_\odot$, $\alpha = 76^\circ - 78^\circ $ so that from eq.(10a),
$$\frac {L_{GW}} {L_{EM}}= 0.54 \bigg (\frac {\beta_6} {\alpha_3} \bigg )^2 \bigg (\frac {R} {10^{8.8} \ cm} \bigg )^6 \bigg (\frac {P} {33 \ s} \bigg )^{-2} \bigg ( \frac {M} {1 \ M_\odot } \bigg )^{-2} \bigg (\frac {B_p} {10^8 \ G} \bigg )^{-2} \bigg ( \frac {<B^2>} {(10^{11}\ G)^2 }\bigg )^2 \bigg (\frac { \sin \alpha } {\sin 77^\circ} \bigg )^2 $$

While for the WD in  RX J0648.0-4418 system [16-18]: 
$ R= 3 \times 10^8$ cm, $ P= 13.184$ s, $M=1.28 \ M_\odot  $.
If one assumes the other parameters to have same values as in the former case, one obtains $ L_{GW}/ L_{EM} \sim 0.02 $ in the case of the WD in RX J0648.0-4418.

Assuming the WD to be in quasi-hydrostatic equilibrium, one may employ the scalar virial theorem (SVT) [20],
$$ 2T + W + 3 \Pi + \mathcal{M}= 0 \ ,\eqno(11)$$
where $T \equiv (\kappa_3\ J^2)/ (2 M\ R^2)$, $W \equiv - (\alpha_3\ G \ M^2)/R $, $  3 \Pi \equiv (\beta_3\ M^{4/3})/ R$ and $\mathcal{M}\equiv \int {(B^2/8\pi) d^3r} $ are the energies  associated with rotation, gravitation, degeneracy pressure and the magnetic field, respectively.

The total energy of the WD is,
$$ E=T + W + 3 \Pi + \mathcal{M} \ .\eqno(12)$$
Eqs.(11) and (12) imply $ E=\ -T$ so that loss of energy incurred because of emission of GWs and EM waves leads to, 
$$\frac {dE} {dt}=\  -\ \frac {dT} {dt}=\ -\ (L_{EM} + L_{GW}) \eqno(13)$$

Hence, from eq.(13), one finds $\frac {dT} {dt} > 0 $, indicating that as the rapidly spinning WD loses energy by radiating GWs as well as EM waves, the rotational kinetic energy of the compact object tends to increase with time. Depending on whether the loss of energy due to radiation causes the WD to  grow or shrink in size will determine whether it slows down or  spins up  (also, see [21] and [22]). 

Furthermore, in the framework of SVT, the rate of change of the WD's period of rotation as it emits GWs and EM waves are  given by, 
$$ \bigg \vert \frac {dP} {dt} \bigg \vert_{GW}= 2.7\times 10^3 \bigg (\frac {\alpha_3 \ T} {0.16\kappa_3 \beta^2_I \ |W|} \bigg )^{2 /3} \bigg (\frac {\beta_6 \delta_M \sin^2 \alpha} {\sqrt{\beta_I}} \bigg )^2 \bigg (\frac {G\ M \ \Omega} {c^3} \bigg )^{5/3} \ $$
 and,
 $$\bigg \vert \frac {dP} {dt} \bigg \vert_{EM}= \frac {5 \pi^2} {P} \frac {B^2_p \ R^4 \ \sin^2 \alpha} {\beta_I \ M \ c^3} \ ,$$
respectively.

\section*{CONCLUSIONS}
Many a surprise is still awaited from GR, particularly, in the strong gravity regime as GR  is a nonlinear dynamical system. Since strong space-time curvatures prevail near compact objects, it is very likely that imprints of  new GR effects will show up in AGN, galactic X-ray source, GRB and neutron star/magnetar studies  involving  data from  sensitive and high resolution telescopes like Astrosat, TMT and SKA.

It will be worth watching out for spin down or spin up that is likely to be exhibited by highly magnetized and rapidly rotating WDs. Such observations would indicate the extent of deformation of WDs that is induced by large magnetic fields. Direct detection of GWs from these sources would require space borne instruments like LISA.

\section* {Acknowledgements}
It is a pleasure to thank Indranil Chattopadhyay and his co-organizers of RETCO-II held at ARIES, Nainital, for creating an  academically stimulating ambience during the conference. 

\section*{References}

 [1] Piotrovich, M. Yu.,  Gnedin, Yu.N.,  Silant'ev, N.A.,   Natsvlishvili, T.M. and   Buliga, S.D., arXiv:1410.1663v2 [astro-ph.HE], and the references therein. 
 
 [2] McHardy, I. M, Koerding, E., Knigge, C., Uttley, P. and Fender, R. P., Nature  444, 730 (2006), and the references therein.

 [3] Penrose, R., Rivista del Nuovo Cimento (1969)
 
 [4] Bhat, M., Dhurandhar, S. and Dadhich, N., J. Astrophys. Astr. 6, 85 (1985)
 
 [5] Itzykson, C. and Zuber, J-B., Quantum Field Theory (McGraw-Hill Inc., 1980)
 
 [6] Hawking, S. W., Commun. Math. Phys. 43, 199 (1975)
 
 [7] Das Gupta, P., arXiv:1505.00541v1 [gr-qc] (to appear in Current Science)
 
 [8] Hawking, S. W., Phys. Rev. D 72, 084013 (2005)
 
 [9] Chen, P., Ong, Y. C. and  Yeom, D., arXiv:1412.8366v2 [gr-qc], and the references therein.
 
 [10] Lasenby, A. N., Dolan, C. J. L., Pritchard, J., Caceres, A. and Dolan, S. R., Phys. Rev. D 72, 105014 (2005)
 
 [11] Schmidt, G. D., et al., Astrophys. J. 595, 1101 (2003) 

 [12]  Kawka, A.,  Vennes,  S.,  Schmidt, G. D.,  Wickramasinghe, D. T. and  Koch, R., Astrophys. J. 654, 499
(2007)

 [13] Liebert, J., et al., Astron. J. 129, 2376 (2005)

 [14] Heyl, J. S.,  Mon.Not.Roy.Astron.Soc. 317, 310 (2000)

 [15] Ikhsanov,  N.R. and Beskrovnaya,  N.G., Astronomy Reports 56, 595 (2012)

 [16] Mereghetti, S.,  Tiengo, A.,  Esposito, P.,  La Palombara, N.,  Israel, G. L.  and  Stella, L., Science 325, 1222 (2009)

 [17] Mereghetti, S., La Palombara, N.,  Tiengo, A.,  Sartore, N., Esposito, P., Israel, G. L. and Stella, L. (arXiv:1304.1653)  

 [18] Mereghetti, S., Proceedings of 13th Marcel Grossman Meeting (MG13), Stockholm, Sweden (2012)(arXiv:1302.4634)

 [19] Das Gupta, P. and Garg, U., in preparation.

 [20] Shapiro, S. L. and  Teukolsky, S. A., Black Holes, White Dwarfs and Neutron Stars (John Wiley and Sons, Inc., 1983)
 
 [21] Shapiro, S. L., Teukolsky, S. A. and Nakamura, T., Astrophys. J. Lett. 357, L17 (1990)

 [22] Boshkayev, K.,  Rueda, J. A.,  Ruffini, R. and Siutsou, I., Astrophys. J. 762, 117 (2013)

\end{document}